\documentstyle[12pt]{article}

\newcommand{\dmuo}{\partial^\mu}
\newcommand{\dmuu}{\partial_\mu}
\newcommand{\pkt}{\; .}
\newcommand{\kma}{\; ,}

\newcommand{\ko}{\; ,}

\newcommand{\intn}[2]{\int\!\frac{{ d}^{#1}\!#2}{(2\pi)^{#1}}}
\newcommand{\inte}{\int\!\frac{{ d}^{3}\!p}{ (2\pi)^{3}2E_0}}
\newcommand{\inteps}{\int\!\frac{{ d}^{n-1}\!p}{ (2\pi)^{n-1}2E_0}}

\newcommand{\re}[1]{{\rm Re }\,#1}
\newcommand{\im}[1]{{\rm Im }\,#1}
\newcommand{\be}{\begin{equation}}
\newcommand{\ee}{\end{equation}}
\newcommand{\bea}{\begin{eqnarray}}
\newcommand{\eea}{\end{eqnarray}}
\newcommand{\eqn}[1]{(\ref{#1})}
\newcommand{\bfx}{{\bf x}}
\newcommand{\bfp}{{\bf p}}
\newcommand{\calf}{{\cal F}}

\newcommand{\cale}{{\cal E}}
\newcommand{\calp}{{\cal P}}
\newcommand{\calh}{{\cal H}}

\newcommand{\call}{{\cal L}}

\newcommand{\calt}{{\cal T}}
\newcommand{\calv}{{\cal V}}
\begin{document}
\begin{titlepage}
\begin{flushright}
DO-TH-99/23\\
December 1999
\end{flushright}
\vspace{20mm}
\begin{center}
{\Large \bf
Renormalization of the nonequilibrium dynamics of fermions
in a flat FRW universe.}

\vspace{10mm}
{\large  J\"urgen Baacke
\footnote{ e-mail:~baacke@physik.uni-dortmund.de}
 and Carsten P\"atzold
\footnote{e-mail:~dr.\_carsten\_paetzold@trinkaus.de}
\vspace{15mm}}

{\large Institut f\"ur Physik, Universit\"at Dortmund} \\
{\large D - 44221 Dortmund , Germany}
\vspace{10mm}

{\bf Abstract}
\end{center}
We derive the renormalized equations of motion and the renormalized
energy-momentum tensor for fermions coupled to a spatially
homogeneous scalar field ( inflaton) in a flat FRW geometry.
The fermion back reaction to the metric and to the inflaton field is 
formulated in one-loop approximation. Having determined the infinite 
counter terms in an $\overline{MS}$ scheme we formulate
the finite terms  in a form suitable for numerical computation.
We comment on the trace anomaly which is inferred from the standard
analysis. We also address the problem of initial singularities
and determine the Bogoliubov transformation by which they are
removed.

\end{titlepage}

\section{Introduction}
The production of fermions during inflation \cite{Inflation}
 was for a long time 
considered to be less interesting than the one of scalar
quanta. The production of scalars coupled to the inflaton
field is characterized, during preheating\cite{KoLiSta,ShtTrBr}, by 
parametric resonance bands
\cite{KoLiSta,BodVHo}, a feature that leads to an abundant production of 
such quanta. Numerous authors have investigated various aspects
of the production of scalar quanta, including the back reaction
to the inflaton field and to the scale parameter
$[5-14]$.
Also, parametric resonance plays an important r\^ole in the analysis of
the late time behavior of these coupled systems
\cite{BdVHSal,Sal}, in the large-$N$ limit.  
For fermions a true parametric resonance cannot develop, as the
fermion number is limited by the Pauli principle. So fermions
were not considered in the inflationary or preheating stage
of inflation but thought to be produced afterwards via the decay of
scalar fields. Recently it
was observed \cite{BHPferm,GreeKo}, that the
 production of fermions
during the preheating stage, via their coupling to the inflaton,
is characterized by 
 resonance-type bands within which the occupation number comes 
near saturation. On the other hand, the production of fermions via 
decay of a boson into an fermion-antifermion pair leads to
a sharp momentum spectrum. 
As a consequence the coherent production of fermions in the 
inflaton background may largely exceed the production rate obtained
via the decay of scalar fields.

This phenomenon has lead several groups to (re)consider the
production of fermionic degrees of freedom during the
inflationary stage. 
 The nonperturbative production of
fermions in an axial background field was considered 
in \cite{MaMa1}.
Very massive fermions may produce characteristic
spikes in the spectrum of primordial perturbations,
leading to observable features in the 
measured CMB anisotropy and in the power spectrum \cite{ChuKoRiTka}.
In supergravity gravitinos 
 will necessarily be a part of the theory, their parameters 
can be related to the scale of supersymmetry breaking. The 
coherent, non thermal production of
gravitinos has been considered 
in Refs. \cite{MaMa2,GiuRiTka,KaKoLivPr}. It may lead to
an exceedingly low reheating temperature and then becomes a problem
for many models of the early universe
\cite{GiuTkaRi,GiuPeRiTka}. This issue still is still open
and being considered by several authors.
One of the problems is the modification of the parametric resonance 
bands by the evolution of the scale parameter,
or by an evolution of the inflaton field modified by the back reaction.

Here we will consider
the renormalization of the fermionic back-reaction to the inflaton field
and to the energy momentum tensor in an FRW geometry, in one-loop
approximation.  
Though this seems a more formal aspect, it is important for 
a reliable numerical computation of these back-reaction effects.
As long as one just considers the number of produced particles
this is not necessary as the adiabatic particle number 
is finite {\sf per se}. For the computation of the
other quantities 
simple normal ordering is not sufficient and
leaves at least logarithmic divergences. 
 
The renormalization of 
nonequilibrium dynamics of fermions in Minkowski space 
has been considered 
by us previously \cite{BHPferm}, but this analysis has to be adapted
to the FRW geometry. In particular dimensional regularization 
is modified by the time dependent scale factor. While the
flat space counter terms still renormalize the divergent loops the
finite terms are different. Furthermore, the structure of the
 energy  momentum tensor in the FRW geometry is modified
as compared to the one in Minkowski space. So several new
renormalization constants and finite parts have to be determined. 
For the case of scalar quantum fluctuations we have already presented
\cite{BHPfrw,BPfrw} our scheme \cite{BHPnumrec} for renormalizing the 
equation of motion for the inflaton and the Friedmann equations.
 The methods used there,
well suited for numerical computations,
can be adapted  to the case fermions. In this sense
the present publication merges our previous work on fermion
systems in Minkowski, and scalar fields in FRW space.
Another viable scheme is the well-known adiabatic subtraction
\cite{PaFu}.
While this has been worked out and numerically implemented
 for the cosmological nonequilibrium
evolution of scalar fields in Ref. \cite{RaHu,HaMoMo}, 
an application to fermions
has not yet been adapted for the requirements of numerical
computation.

Of course not all the results presented in this publication are 
new. The renormalization of the energy momentum tensor of 
free fermions in a FRW background
has been considered long ago, see e.g. \cite{Davies:1977,
Bunch:1979,BiDa}; 
other regularization schemes have been 
discussed, e.g., in Refs. \cite{MaMo1,MaMo2};
the one-loop  renormalization
of a Yukawa theory is of course standard. However,  while the
structure of the renormalization constants is known, we have to
rederive them here within the nonequilibrium formalism.
An important point is their independence of the initial conditions.
Furthermore, the remaining finite parts have to be written in a form
suitable for numerical computation. As usual, the formulation of the
finite parts is the most cumbersome part of renormalization,
and this has not been discussed before.

The plan of the paper is as follows:
in section 2 we recall the basic formulae for a flat Friedmann
universe, with $3+\epsilon$ space dimensions.
In section 3 we embed into this geometry the quantum field theory describing
massive fermions, Yukawa-coupled to the scalar inflaton field. 
The energy momentum tensor for the classical field
and the fermionic quantum fluctuations is given in section 4.
In section 5 we determine the renormalization constants, and thereby the
precise formulae for the finite parts. We also include the trace anomaly
that cannot be derived from considering  the FRW metric alone.
We briefly digress  on the question of initial conditions in 
section 6, conclusions are presented in section 7.


\section{FRW cosmology}
\setcounter{equation}{0}
We consider the Friedmann-Robertson-Walker metric
with curvature parameter $k=0$, i. e. a spatially isotropic and flat
space-time.
The line-element is given in this case by
\begin{equation} 
\label{line}
{d}s^2={d}t^2-a^2(t){d}{\bfx} ^2\; .
\end{equation}
The time evolution of the cosmic scale factor $a(t)$ is governed
by Einstein's field equation
\be\label{Einfield}
(1+\delta Z) G_{\mu\nu}+
\delta\alpha\,^{(1)}H\,_{\mu\nu}+\delta\beta\,^{(2)}H
\,_{\mu\nu}
+\delta\gamma\,^{}H\,_{\mu\nu}+\delta\Lambda g\,_{\mu\nu}=-\kappa\langle
T_{\mu\nu}
\rangle
\ee
with $\kappa= 8\pi G$. 

The Einstein curvature tensor $G_{\mu\nu}$ is given by
\begin{equation} 
G_{\mu\nu}=R_{\mu\nu}-\textstyle{\frac 1 2} g_{\mu\nu}R\; .
\end{equation}
The Ricci tensor and the Ricci scalar are defined as
\begin{eqnarray}
R_{\mu\nu}&=&R^{\lambda}_{\mu\nu\lambda}\; ,\\
R&=&g^{\mu\nu}R_{\mu\nu}\; ,
\end{eqnarray}
where
\begin{equation} 
R^{\lambda}_{\alpha\beta\gamma}=
\partial_{\gamma}\Gamma^{\lambda}_{\alpha\beta}
-\partial_{\alpha}\Gamma^{\lambda}_{\gamma\beta}
+\Gamma^{\lambda}_{\gamma\sigma}
\Gamma^{\sigma}_{\alpha\beta}
-\Gamma^{\lambda}_{\alpha\sigma}\Gamma^{\sigma}_{\gamma\beta}\; .
\end{equation}
In FRW geometry with $n-1$ space dimensions, as needed for
dimensional regularization, the relevant quantities
are given by
\bea\nonumber
R_{tt}&=&(n-1)\frac{\ddot a}{a}\\
&&\label{Ricci_ndim}\\ \nonumber
R^\mu_\mu= R&=&2(n-1)\frac{\ddot a}{a}+
(n-1)(n-2)\left(\frac{\dot a}{a}\right)^2
\pkt
\eea
It is customary to introduce the Hubble parameter
$\dot a/a$ and to use $R= (n-1)(2 \dot H + n H^2)$.
The tensors  $\,^{(1)}H_{\mu\nu}, \,^{(2)}H_{\mu\nu}$,
and $H_{\mu\nu}$ arise from the variation
of terms proportional to $R^2,R\,^{\alpha\beta}R_{\alpha\beta}$,
and $R\,^{\alpha\beta\gamma\delta}R_{\alpha\beta\gamma\delta}$ 
in the Hilbert-Einstein action.
Their general definition is given, e.g.,  in \cite{BiDa}. Though 
the fermion loop integrals  will  not produce divergences
proportional to these tensors we will need them as they 
produce the trace anomaly. The explicit expressions in FRW geometry 
with $n-1$ space dimensions (see, e.g. \cite{BPfrw}) are presented
 in Appendix A.

The Friedmann equations are obtained by inserting the explicit 
expressions for the geometrical tensors and for the energy-momentum
tensor into the Einstein equations. As this is done after renormalization
we can set $n=4$. Then  
\bea
3 H^2 &=& \kappa \cale_{\rm ren} \\
R &=&  \kappa \calt_{\rm ren}
\pkt
\eea
Here $\cale = T_{tt}$ and $\calt = T^\mu_\mu$.
For the numerical evolution it is sufficient to consider just
the first equation. Then the second one is fulfilled as a consequence
of covariant energy conservation
\be
\dot \cale + H (4 \cale - \calt)=0
\pkt
\ee 
 

\section{Lagrangian and equations of motion}
\setcounter{equation}{0}
We consider the production of fermions by a scalar field
$\Phi$ 
 in a Heisenberg state in which the scalar field 
has a time-dependent, but spatially homogeneous expectation value
\be
\phi(t) = \langle \Phi(t) \rangle
\ee
The Lagrangian density of the model in
Minkowski space is given by
\bea \nonumber
\call& =& \frac{1+\delta Z}{2}\dmuu\Phi\dmuo\Phi - \calv(\Phi)
-\frac{\xi+\delta \xi}{2} R \Phi^2
- (\zeta+\delta \zeta) R \Phi-
\\ \label{baslag}
&&+\bar\psi\left(i\gamma^\mu\dmuu -m_f-g\Phi\right)\psi
\eea
where 
\bea \nonumber
\calv(\Phi)&=&
(\sigma+\delta \sigma) \Phi
+\frac{M^2+\delta M^2}{2}\Phi^2 \\  
&&+\frac{\kappa+\delta \kappa}{6} \Phi^3
+\frac{\lambda+\delta\lambda}{24}\Phi^4 
\eea
In addition to the scalar  mass $M$,
the quartic self-coupling $\lambda$  and the
conformal coupling $\xi$ we have introduced a tadpole coupling
$\sigma$ , a trilinear coupling $\kappa$ ,
and an additional  coupling
to the curvature tensor $\zeta R \Phi$.
For $m_f \neq 0$ such terms are generated by the fermion
loop and at least the counterterms are needed for the purpose of
renormalization. Alternatively one may generate
the fermion mass by a shift in $\Phi$, starting with a massless fermion;
then the divergent terms with odd powers of $\Phi$  are generated 
from the terms with
even powers of $\Phi$ obtained in the massless case. 
We have written the counter terms explicitly from the outset
though their form is obvious. For the wave function
renormalization this is necessary:
if we use conformal time and 
conformal rescaling of the
fields in $n=4-\epsilon$ dimensions, then part of the
kinetic term, including the divergent wave function renormalization,
reappears as part of the conformal coupling (see below).
  
The action in curved space is obtained
in the usual way by introducing covariant derivatives
and the covariant measure, we have anticipated already the
couplings to the curvature tensor. We will implement dimensional
 regularization in the way of using $n-1$ space components,
so we will have to use
FRW geometry with $n-1=3-\epsilon$ space components, as well. 

The equation of motion of the condensate is given by
\bea \label{phidglt}
(1+\delta Z)\left[\ddot\phi+(n-1)\frac{\dot a}{a}\dot \phi\right]
 +\frac{d\calv(\phi)}{d\phi}+ 
(\zeta+\delta \zeta) R + 
 (\xi +\delta\xi) R \phi(t)
\\ \nonumber
+ \frac{\lambda +\delta\lambda}{2}\langle\eta^2\rangle
+g\langle\overline{\psi}\psi\rangle=0\pkt
\eea
where the back reaction of the fermion field is given, in
1-loop and Hartree approximation by the term
$g\langle\overline{\psi}\psi\rangle$, the back-reaction of the
scalar field may be included in one-loop, Hartree, or
large-$N$ approximation. 
The renormalization of the back-reaction of the scalar fluctuations 
has been discussed recently within our scheme
in Refs. \cite{BHPfrw,BPfrw}, and in the adiabatic subtraction scheme
in Ref. \cite{RaHu,HaMoMo}. So we do not consider these fluctuations here.

It is convenient to introduce conformal time
\be
dt=a(\tau)d\tau
\ee
and to  rescale the field $\phi$ via
\be
\phi(\tau)=a^{-1+\epsilon/2}(\tau)\tilde \phi(\tau) 
\pkt
\ee
The derivative with respect to conformal time will be denoted
\be
f'(\tau)\equiv \frac{\partial}{\partial \tau}f (\tau)
= \partial_\tau f (\tau)
\pkt
\ee 
The classical equation of motion now takes the form
\bea \nonumber
&&(1+\delta Z) \tilde \phi''
+(\zeta+\delta\zeta) a^{3-\epsilon/2} R +
a^2 
\left[\xi+\delta\xi-(1+\delta Z)\frac{n-2}{4(n-1)} R\right]
\tilde\phi 
\\ 
&&a^{4-\epsilon}\frac{d\calv_\epsilon(\tilde\phi)}{d\tilde\phi}
+\frac{\lambda+\delta \lambda}{2}a^\epsilon\tilde \phi\langle\eta^2\rangle
+ga^{3-\epsilon/2}\langle\overline{\psi}\psi\rangle=0\kma
\eea
where we have introduced the rescaled potential
\bea\nonumber
\calv_\epsilon(\tilde \phi)&=&
(\sigma+\delta\sigma)a^{-1+\epsilon/2}\tilde\phi
+\frac{M^2+\delta M^2}{2}a^{-2+\epsilon}\tilde\phi^2
\\&&\label{veps}
+\frac{\kappa+\delta\kappa}{6}a^{-3+3\epsilon/2}\tilde\phi^3
+\frac{\lambda+\delta\lambda}{24}a^{-4+2\epsilon}\tilde\phi^4
\eea
The covariant derivative in the Dirac equation is obtained
using the $n$-bein formalism (see Appendix B). One obtains
\be
i\gamma^0\left(\partial_0+\frac{n-1}{2}\frac{\dot a}{a}\right)\psi
+ \left(\frac{i}{a}\gamma^k\nabla_k -m_f-
g a^{\epsilon/2}\tilde \phi\right)\psi=0
\pkt
\ee
The extra term in the first parenthesis can be eliminated
 by defining
\be
\psi=a^{-(n-1)/2}\tilde\psi
\; ; \ee
furthermore, we introduce conformal time
and  obtain
\be
\left[i\partial_\tau -\tilde{\cal H}(\tau)\right]
\tilde\psi(\tau,\bfx)=0
\pkt
\ee
with
\be
\tilde{\cal H}(\tau)=-i\mbox{\boldmath$\alpha$\unboldmath}
{\bf \nabla} + \left[a(\tau) m_f 
+g a^{\epsilon/2}(\tau)\tilde\phi(\tau)\right]\beta
\pkt \ee
Note that $\tilde\calh$ is rescaled with respect to the Hamiltonian
in Minkowski space with a factor $a(\tau)$.
So the Dirac equation takes its conventional form, with an effective
time-dependent mass
\be \label{meff}
m(\tau)= a(\tau) m_f 
+g a^{\epsilon/2}(\tau)\tilde\phi(\tau)
\pkt \ee
On account of the spatial homogeneity of the condensate field it
is suitable to expand the Dirac field as
\be
\tilde\psi(\tau,{\bf x})=\sum_s\inte\left[
b_{\bfp,s}U_{\bfp,s}(\tau)+d^\dagger_{-\bfp,s}{V}_{-\bfp,s}(\tau)
\right]e^{+i\bfp\cdot\bfx}\kma
\ee
with the time independent creation and 
annhiliation operators for quanta, whose mass $m_0=m(0)$
 is determined by the initial 
state. $E_0$ is the corresponding energy $\sqrt{m_0^2+\bfp^2}$.
The creation and annihilation operators
satisfy the standard anti-commutation relations
\bea
\{b_{\bfp,s},b^\dagger_{\bfp',s'}\}&=&2 E_0(2\pi)^3\delta^3(\bfp-\bfp')
\delta_{ss'}\kma\\
\{d_{\bfp,s},d^\dagger_{\bfp',s'}\}&=&2 E_0(2\pi)^3\delta^3(\bfp-\bfp')
\delta_{ss'}
\pkt
\eea
For the positive and negative energy solutions we make the usual  ansatz
\be
U_{\bfp,s}(\tau)=N_0\left[i\partial_{\tau}+ 
\tilde{\cal H}_\bfp(\tau)\right]f_p(\tau)
\left(\begin{array}{c}
\chi_s \\ 0
\end{array}\right)
\ee
and
\be
V_{\bfp,s}(\tau)=N_0\left[i\partial_{\tau}+ 
\tilde{\cal H}_{-\bfp}(\tau)\right]g_p(\tau)
\left(\begin{array}{c}
0 \\ \chi_s
\end{array}\right)\kma
\ee
with the Fourier-transformed Hamiltonian
\be
\tilde{\cal H}_\bfp(\tau)=\mbox{\boldmath $\alpha p$\unboldmath} +m(t)\beta
\pkt \ee
For the two-spinors $\chi_s$ we use helicity eigenstates, i.e.,
\be
\mbox{\boldmath$ \hat p\sigma$\unboldmath} \chi_\pm=\pm \chi_\pm
\pkt
\ee
The mode functions $f_p$ and $g_p$ depend only on $p=|\bfp|$;
they obey the second order differential equations
\bea\label{fsec}
\left[
\frac{d^2}{d\tau^2}-im'(\tau)+ p^2+m^2(\tau)
\right]f_p(\tau)&=&0\kma \\\label{gsec}
\left[
\frac{d^2}{d\tau^2}+im'(\tau)+p^2+m^2(\tau)
\right]g_p(\tau)&=&0\pkt
\eea
If the fermion Fock space is based on the conformal vacuum state
the modes start - in conformal time - as if the mass 
was independent of $\tau$ for $\tau \le 0$.
The  mode functions, which would be plane waves
for $t\le 0$, then satisfy the initial conditions
\bea\label{fginit}
f_p(0)=1&,& \dot{f}_p(0)=-iE_0\kma\\
g_p(0)=1&,& \dot{g}_p(0)=iE_0\kma
\eea
so that $g(\tau)=f^*(\tau)$.
The fermion condensate occurs in the equation of
motion as $ga^{3-\epsilon/2}\langle\bar\psi\psi\rangle$,
the rescaling of the Dirac field changes this to
$ga^{\epsilon/2}\langle \bar{\tilde \psi }\tilde\psi\rangle$.
We define a fluctuation integral as
\bea 
\calf(\tau)&=& \langle\bar{\tilde\psi}\tilde\psi\rangle
=\sum_s\inte\overline{V}_{-\bfp,s}(t)V_{-\bfp,s}(t)
\nonumber\\ \label{flucint}
&=&-2\inteps\left\{
2E_0-\frac{2\bfp^2}{E_0+m_0 }|f_p|^2
\right\}\pkt
\eea
We have written the mode integral in dimensionally regulated form.
This corresponds to the usual prescription of introducing
dimensional regularization `after taking the Dirac traces'.
Indeed the mode integral is a trace of a Green function.


\section{The energy-momentum tensor}
\setcounter{equation}{0}
The energy-momentum tensor for a spatially isotropic background field
and of the quantum fluctuations generated by this field is
diagonal and of the form
\be
T^{\mu\nu} ={\rm diag}\left(\cale,\calp,\calp,\calp\right)
\pkt 
\ee
From the Lagrangian \eqn{baslag} we derive for the
energy density of the condensate or background field
\bea
\cale_{\rm cond}&=&
\frac{1+\delta Z}{2}a^{-4+\epsilon}(\tilde\phi')^2
+\calv_\epsilon(\tilde\phi)
\\&&\nonumber
+2(n-1)\left[\xi+\delta\xi-\frac{(n-2)(1+\delta Z)}{4n-4}\right]
a^{-2+\epsilon}
\left(Ha^{-1}\tilde\phi\tilde\phi'-\frac{n-2}{4}H^2\tilde\phi^2
\right) \\ &&\nonumber
+2(\zeta+\delta \zeta)(n-1)a^{-2+\epsilon/2}
H\tilde\phi'
+\frac{\delta Z_G}{\kappa}G_{tt} 
+\frac{\delta\Lambda}{\kappa}
\kma\eea
where $\calv_\epsilon(\tilde \phi)$
has been defined above, in Eq. \eqn{veps}.

We have included the counter terms proportional to $G_{tt}$
and $g_{tt}$ which appear on the left
hand side of the Einstein field equations. They are needed, as
the other counter terms, in order to compensate the divergences
arising in the fluctuation part of $T_{tt}$. 
The higher curvature counter terms on the left hand side of
the Einstein equation \eqn{Einfield} are not needed
for infinite renormalizations, they will play
a r\^ole, however, in the discussion of the trace anomaly 
at the end of section 5.

Instead of the pressure we consider the trace of the
energy-momentum tensor. We denote it by
$\calt=\cale-(n-1)\calp$. For the background field we find
\bea\nonumber
\calt_{\rm cond}&=&
n\calv_\epsilon(\tilde\phi)
\\&&\nonumber
+2(n-1)\left[\xi+\delta\xi-\frac{(n-2)(1+\delta Z)}{4n-4}\right]
a^{-2+\epsilon}
\left(a^{-1}\tilde\phi'-\frac{n-2}{2}H\tilde\phi
\right)^2 \\ &&\nonumber
+2(n-1)(\zeta+\delta \zeta)a^{-3+\epsilon/2}
\left[\tilde\phi''+\frac{(n-2)Ra^2}{n-1}\tilde\phi\right]
+\frac{\delta Z_{\rm G}}{\kappa}G^\mu_\mu
+n \frac{\delta\Lambda}{\kappa}
\kma\eea
including again counter terms from the left hand side
of the Einstein equations. 

The fluctuation parts of the energy-momentum tensor
are given, after conformal rescaling, by
\bea \nonumber
{\cal E}_{\rm fl}(\tau)&=&a^{-4+\epsilon}\langle
\overline{\tilde\psi}\left(\beta \tilde{\cal H}_p\right)\tilde\psi\rangle
\\ \label{e_int} &=&a^{-4+\epsilon}\sum_s\inte\overline{V}_{-\bfp,s}(\tau)
\left(\beta\tilde{\cal H}_p\right) V_{-\bfp,s}(\tau)
\\\nonumber
&=&2a^{-4+\epsilon}\inteps\left\{i\left[E_0-m_0)\right]
\left(f_p f^{*'}_p-f'_pf^*_p\right)-2E_0 m(\tau)\right\}
\pkt\eea
for the energy density and by
\bea \label{p_int}
{\cal P}_{\rm fl}(\tau)&=&
\frac{a^{-4+\epsilon}}{n-1}\langle
\overline{\tilde\psi}\mbox{\boldmath $\gamma p$\unboldmath} 
\tilde\psi\rangle
\\ \nonumber
&=&\frac{a^{-4+\epsilon}}{n-1} \sum_s\inte\overline{V}_{-\bfp,s}(\tau)
\mbox{\boldmath $\gamma p$\unboldmath} V_{-\bfp,s}(\tau)
\eea
for the pressure. The fluctuation part of the energy-momentum tensor
is given by
\be
\left(T^\mu_\mu\right)_{\rm fl} = \cale- (n-1)\calp
\pkt
\ee
This results in
\bea \nonumber
\calt_{\rm fl}(\tau)&=&a^{-4+\epsilon}\sum_s\inte\overline{V}_{-\bfp,s}(\tau)
\left(\beta\tilde{\cal H}_p-\mbox{\boldmath $\gamma p$\unboldmath}\right)
 V_{-\bfp,s}(\tau)
\\&&\label{Tfluct} = a^{-4+\epsilon}\sum_s\inte\overline{V}_{-\bfp,s}(\tau)
m(\tau) V_{-\bfp,s}(\tau)
\\ \nonumber
& &= m(\tau)a^{-4+\epsilon}\calf(\tau)
\pkt
\eea
So the trace can be computed by multiplying the fluctuation
integral $\calf$ by the time-dependent mass.


\section{Renormalization}
\setcounter{equation}{0}
In order to develop the framework for
renormalizing the one-loop equations,
we write the equation of motion for the mode 
functions, Eq. \eqn{fsec}, in the form
\be  \label{f_diff_eq}
\left[ \frac{d^2}{ d\tau^2}+
E_0^2\right]f_p(\tau)=-V(\tau)f_p(\tau)\kma
\ee
with
\be \label{pot_def}
V(\tau)=m^2(\tau)-m(0)^2-i\left[m'(\tau)
-m'(0)\right]\pkt
\ee
Using the initial conditions \eqn{fginit}
this equation can be recast into the form of
an integral equation:
\be \label{f_int_eq}
f_p(\tau)=e^{-iE_0 \tau}
-\frac{1}{E_0}\int_0^\tau d\tau' \sin[E_0(\tau-\tau')]V(\tau')f_p(\tau')
\pkt
\ee
Using this integral equation, the mode functions may be expanded
with respect to the potential $V(\tau)$.
We split off the zeroth order (plane wave) contribution and
an oscillating phase factor by writing
\be \label{h_def}
f_p(\tau)=e^{-i E_0 \tau}[1+h_p(\tau)] \pkt
\ee
The differential and integral equations
satisfied by the function  $h_p(\tau)$ can easily be derived
 from Eqs. \eqn{f_diff_eq}
and \eqn{f_int_eq}, respectively. 
It may be decomposed as
\be \label{h_excl}
h_p(\tau)=\sum_{n=1}^\infty h_p^{(n)}( \tau),
\ee
where $h_p^{(n)}(\tau)$ is of $n$'th order in $V(\tau)$; 
we define further the inclusive sums
\be\label{h_incl}
h_p^{\overline{(n)}}=\sum_{m=n}^\infty h_p^{(m)}
\pkt \ee
The truncated mode functions $h_p^{(n)}(\tau)$ and  $h_p^{\overline{(n)}}$
satisfy a recursive set of differential and integral equations 
derived from
Eqs. \eqn{f_diff_eq} and \eqn{f_int_eq}; these can be used to
compute them numerically and analytically. 
Moreover the expansions
\eqn{h_excl} and \eqn{h_incl} are expansions with respect
to $1/E_0$, we have used them \cite{BHPferm} 
in order to single out the leading
divergent contributions in the various mode integrals.

The integrand of the 
fluctuation integral $\calf$, see Eq. \eqn{flucint},
can be written as 
\bea \nonumber
1-\left(1-\frac{m_0}{E_0}\right)|f_p(\tau)|^2&=&\frac{m_0}{E_0}
-\left(1-\frac{m_0}{E_0}\right)\left[ 2 \re h_p(\tau)+ |h_p(\tau)|^2\right]
\\ \nonumber 
&=&\frac{m(\tau)}{E_0}
-\frac{m''(\tau)}{4(E_0)^3}-
\frac{m^3(\tau)}{2(E_0)^3}+
\frac{m(\tau)m_0^2}{2(E_0)^3}
\nonumber \\ && \label{flucintexp} 
+\frac{m''(0)}{4(E_0)^3}\cos \left(2E_0 \tau\right)+
K_{\rm F}(p,\tau)
\pkt
\eea
The first terms on the right hand side lead to divergent 
or singular momentum integrals. The function $K_{\rm F}(\tau)$ 
can be considered being defined by this
equation, i.e. by subtraction of the remainder of the right
hand side from the left hand side.
 It behaves as $(E_0)^{-4}$ and its momentum integral
is finite. While other authors 
\cite{HaMoMo} use the cutoff independence
with scale dependent couplings in order to check
their numerical scheme, we use this asymptotic behavior as
a numerical cross check.
If defining $K_{\rm F}(\tau) $ by subtraction is numerically
precarious, it can also be computed directly, using the
reduced mode functions  $h_p^{(n)}(\tau)$ and  $h_p^{\overline{(n)}}$.
 This is discussed in \cite{BHPferm}. For fermions these expressions 
get rather lengthy, we do not present them here. Using the truncated
mode functions instead of subtracting plainly the divergent
parts from the integrand constitutes another numerical and analytical
cross-check.

We decompose the fluctuation integral as
\be\label{Fdecomp}
\calf(\tau)=\calf_{\rm div}(\tau)+\calf_{\rm sing}(\tau)+
\calf_{\rm fin}(\tau)
\kma \ee
with
\bea
\calf_{\rm div}&=&-2\intn{n-1}{p}\left\{\frac{m(\tau)}{E_0}
-\frac{ m'' (\tau)}{4(E_0)^3}-\frac{m^3(\tau)}{2(E_0)^3}+
\frac{m(\tau)m_0^2}{2(E_0)^3}
\right\}\kma\\  \label{fsing}
\calf_{\rm sing}&=&-2\intn{3}{p}\left\{-\frac{ m'(0)}{2(E_0)^2}
\sin(2E_0\tau) +
\frac{m''(0)}{4(E_0)^3}\cos (2E_0\tau)
\right\}\kma\\
\calf_{\rm fin}&=&-2\intn{3}{p} K_{\rm F}(p,\tau)
\pkt\eea
We have dropped dimensional regularization for the 
singular and finite parts. They are to be 
evaluated at $n-4=\epsilon=0$. 
The divergent part has to be compensated by suitable counter terms.
The   part $\calf_{\rm sing}$  displays a
singularity in $\tau$ at $\tau=0$, a phenomenon related to the initial
conditions. One gets rid of the problem by a Bogoliubov
transformation \cite{CooMoinit,BHPinit}, for 
fermions its explicit form has been derived in \cite{BHPinit,BBdVferm}.  
We derive the form of this Bogoliubov transform for the case
under consideration in section 6.

The fluctuation parts of the energy momentum tensor can be
analyzed in a similar way.
The integrand of the energy density $\cale_{\rm fl}$ can be expanded as
\bea \nonumber
&&\frac{i}{2}\left(1-\frac{m_0}{E_0}\right)\left(
f_p f^{*'}_p- f'_p f_p^*\right)-m(\tau)\\ \nonumber
&&=-(E_0-m_0)\left \{ 1+ 2 \re h_p+|h_p|^2
-\frac{1}{E_0}\im\left[  h'_p\left(1+h_p^*\right)\right]\right \}
-m(\tau)\\ \nonumber &&=
-E_0-\frac{m^2(\tau)}{2E_0}+\frac{m^2_0}{2E_0}+
\frac{m'^2(\tau)}{8(E_0)^3}+\frac{m^4(\tau)}{8(E_0)^3}
 \nonumber \\ && 
\hspace{5mm}+\frac{m^4_0}{8(E_0)^3}-\frac{m^2(\tau)m_0^2}{4(E_0)^3}
+K_{\rm E}(p,\tau)\pkt
\eea
Again $K_{\rm E}(p,\tau)$ is defined by this equation and it behaves
as $(E_0)^{-4}$ as $E_0 \to \infty$. There is no cosine term here
and, therefore, no singular contribution.
So 
\be
\cale_{\rm fl}(\tau)=\cale_{\rm fl, div}(\tau)+\cale_{\rm fl, fin}(\tau)
\kma \ee
with
\bea
\cale_{\rm div}&=&2a^{-4+\epsilon}\intn{n-1}{p}
\left\{-E_0-\frac{m^2(\tau)}{2E_0}+\frac{m^2_0}{2E_0}+
\frac{m'^2(\tau)}{8(E_0)^3}+\frac{m^4(\tau)}{8(E_0)^3}
\right. \nonumber \\ && \left.
+\frac{m^4_0}{8(E_0)^3}-\frac{m^2(\tau)m^2_0}{4(E_0)^3} \right\}
\kma\\ 
\cale_{\rm fin}&=&2a^{-4}\intn{3}{p}K_{\rm E}(p,\tau)
\pkt\eea
The contribution of the fluctuations to the trace
is particularly simple: as explained in the previous section,
it is proportionla to the fluctuation integral.
So, $\calt_{\rm fl}(\tau)$ can be decomposed as
\be
\calt_{\rm fl}(\tau)=
\calt_{\rm fl, div}(\tau)+\calt_{\rm fl, sing}(\tau)
+\calt_{\rm fl, fin}(\tau)
\kma \ee
where the three expressions on the right hand side are obtained
from those of Eq. \eqn{Fdecomp} by multiplying them with 
$m(\tau)a^{-4+\epsilon}$.
For discussing the renormalization we need in particular the divergent 
part
\be
\calt_{\rm fl, div}
=a^{-4+\epsilon}
\intn{n-1}{p}\left\{\frac{m^2(\tau)}{E_0}
-\frac{m(\tau)m''(\tau)}{4(E_0)^3}-
\frac{m^4(\tau)}{2(E_0)^3}+
\frac{m^2(\tau)m_0^2}{2(E_0)^3}\right\}
\ee
The divergent terms $\calf_{\rm div},\cale_{\rm div}$,
 and $\calt_{\rm div}$ are proportional to local terms in
$\phi(\tau)$ and its derivatives. These can be absorbed in the usual
way by introducing the appropriate counter terms into the Lagrangian
and into the energy-momentum tensor.

The divergent parts of the fluctuation integral
can be evaluated, e.g., using dimensional regularization.
One finds
\be \label{Fdivex}
\calf_{\rm div}=2m''(\tau)L_0
+4m^3(\tau)L_0
+\frac{m(\tau)m_0^2}{4\pi^2}\kma
\ee
with the abbreviation \footnote{\label{Lvar}
This definition and the one of $L$ and $L_f$  below deviate from
those in our previous work \cite{BHPferm,BPfrw} by inclusion of 
the factor $1/16\pi^2$.}
\be
L_0= \frac{1}{16\pi^2}\left\{\frac{2}{\epsilon}+
\ln{\frac{4\pi\mu^2a^2(\tau)}{m_0^2}}-\gamma\pkt\right\}
\ee  
The scale $\mu$ is introduced as usual by 'adjusting the dimension'
of the various couplings, we have not displayed this rescaling
explicitly.
The initial mass only appears in the logarithm of $L_0$ and in the
last term of \eqn{Fdivex} which is finite. Therefore,
as already found for the scalar fluctuations
\cite{BHPnumrec}, the dependence 
on the initial mass $m_0=m(0)$ can be absorbed into finite terms,
$\Delta Z,\Delta \sigma,\dots$. In order to appreciate
this independence of the initial mass it should be noticed that
the individual divergent terms appear with prefactors up to
$m_0^3$, and up to $m_0^4$ in the energy momentum tensor.
If one uses a three-dimensional 'computer' cutoff the necessary
cancellations do not occur.
 
We expand \eqn{Fdivex} using the explicit expression \eqn{meff} 
for $m(\tau)$. This leads to rather lengthy expressions,
in particular we have to keep terms proportional
to $n-4$, as these become finite terms when 
multiplied with the $1/\epsilon$ of $L_0$. A further source
of finite terms is the product $L_0 a^{\epsilon/2}$ which appears
when $g a^{\epsilon/2} \calf$ is evaluated in  the equation of motion.
 We have used
{\sc maple} for computing the divergent terms and their finite 
remnants in the equation of motion. The second derivative term contains
terms proportional to $a''$, and therefore to the curvature scalar
$R$. So the conformal couplings $\zeta$ and $\xi$ are needed
already in order to renormalize the equation of motion. 
In flat Minkowski space these couplings only appear in the pressure.

Applying an $\overline{MS}$ prescription,
the infinite renormalizations become
\bea \nonumber
\delta Z &=-2 g^2 L & \delta \sigma= -4 g m_f^3 L   
\\ \nonumber
\delta M^2 &=-12 m_f^2 g^2 L   & \delta \kappa = -24 m_f g^3
\\
\delta \lambda &= -24 g^4 L  & \delta \zeta = -\frac 1 3 m_f L 
\\ \nonumber 
\delta \xi &=- \frac 1 3 g^2 L  &
\eea
with 
\be \label{Ldef}
L= \frac{1}{16\pi^2}\left\{\frac{2}{\epsilon}+
\ln{\frac{4\pi\mu^2}{M^2}}-\gamma\pkt\right\}
\ee
We use a scale factor $M$ instead of the fermion mass, as one may
want to consider the case $m_f=0$.
The renormalization of the conformal coupling $\delta\xi$ is due
entirely to the appearance of $\delta Z$ in the conformal 
coupling term. 

The renormalized equation of motion is obtained by replacing
the infinite parts of the counter terms by their finite remnants, and
by including some additional terms.
The finite remnants of the counter terms are
\bea \nonumber
\Delta Z =2 g^2 L_f && \Delta \sigma= 4 g m_f^3 L_f   
\\ \nonumber
\Delta M^2 =12 m_f^2 g^2 L_f  &&  \Delta \kappa = 24 m_f g^3 L_f
\\ \label{finitect1}
\Delta \lambda = 24 g^4 L_f && \Delta \zeta = \frac 1 3 g m_f L_f  
\\ \nonumber 
\Delta \xi = \frac 1 3 g^2 L_f  &&
\eea
with 
\be
L_f= \frac{1}{16\pi^2}
\ln{\frac{M^2 a^2(\tau)}{m_0^2}}
\pkt\ee
These finite parts are {\em time-dependent}, the
term  $\ln a(\tau)$ arises from
the conformal scale factors via the product $(2/\epsilon)a^{\epsilon/2}$.
Conceptually they are part of the remaining nonlocal 
finite fluctuation integrals,
though they can be written in local form.

The renormalized equation of motion, omitting the scalar back-reaction,
reads
\bea \nonumber
&&(1+\Delta Z) \tilde \phi''
+(\zeta+\Delta\zeta) a^3 R +
a^2 
\left[\xi-\frac{1}{6}\right] R \tilde\phi 
+a^4\frac{d\calv_{\rm ren}(\tilde\phi)}{d\tilde\phi}
\\ \label{eqmren} &&
+\frac{g}{16 \pi^2}\left(\frac{2}{9}a^2mR+4m_0^2m+4aHm'
-2a^2H^2m\right) \\ \nonumber
&&+g\left(\calf_{\rm sing}(\tau)+
\calf_{\rm fin}(\tau)\right)
=0\kma
\eea
where we have introduced the renormalized potential
\bea\nonumber
\calv_{\rm ren}(\tilde \phi)&=&
(\sigma+\Delta\sigma)a^{-1}\tilde\phi
+\frac{M^2+\Delta M^2}{2}a^{-2}\tilde\phi^2
\\&&\label{vren}
+\frac{\kappa+\Delta\kappa}{6}a^{-3}\tilde\phi^3
+\frac{\lambda+\Delta\lambda}{24}a^{-4}\tilde\phi^4
\eea
that we will find again in the renormalized stress-energy
tensor.
We note that $\Delta \xi $ and $\Delta Z$ have cancelled in the
conformal coupling term; however, even if $\xi=1/6$ an additional
conformal coupling $\propto g^2$ is introduced by the first term
in the parenthesis in Eq. \eqn{eqmren}. It does not vanish even  for $m_f =0$. 
The cosmological constant counter term has not yet been fixed
as it does not appear in the derivative of the potential.
It will be determined below. 
While the equation of motion is now finite for $\tau \neq 0$ it still 
contains $\calf_{\rm sing}(\tau)$ that is
 singular at $\tau=0$.

The divergent parts of the energy density give, 
after dimensional regularization
\be
{\cale}_{\rm fl, div}=a^{-4+\epsilon}\left\{m'^2(\tau)L_0+
m^4(\tau) L_0+\frac{m^4(0)}{32\pi^2}
+\frac{m^2(\tau)m^2_0}{8\pi^2}\right\}\pkt
\ee
We again expand this, using the explicit expression for 
$m(\tau)$. In addition to the divergent terms found in the
equation of motion we find terms that have to be
compensated by the cosmological constant counter term and
by the wave function renormalization of the
gravitation field.
This fixes these couplings to
\bea \label{einsteinren}
\delta \Lambda &=& -\kappa m_f^4 L \\
\delta Z_G &=& \kappa \frac{m_f^2}{3} L 
\kma
\eea
again independent of the initial condition.
The renormalized energy density is given by
\bea\nonumber
\cale_{\rm ren}&=& 
\frac{1+\Delta Z}{2}a^{-4}(\tilde\phi')^2
+\calv_\epsilon(\tilde\phi)+6(\zeta+\Delta \zeta)a^{-2}H\tilde\phi'
\\&&
+6\left[\xi-\frac{1}{6}\right]
a^{-2}
\left(Ha^{-1}\tilde\phi\tilde\phi'-\frac{1}{2}H^2\tilde\phi^2
\right)
\\ && \nonumber
+\frac{a^{-4}}{96\pi^2}\left(3m_0^4+12m_0^2m^2+8aHmm'+2a^2H^2m^2\right)
\\ \nonumber
&&+\frac{\Delta Z_G}{\kappa}G_{tt} +\frac{\Delta \Lambda}
{\kappa}
+\cale_{\rm fin}
\kma\eea
where we have introduced the finite renormalizations 
\bea \label{einsteinfin}
\Delta \Lambda &=& \kappa m_f^4 L_f \\
\Delta Z_G &=& -\kappa \frac{m_f^2}{3} L_f 
\pkt
\eea
Though these terms have a local form they need not be considered
as renormalizations of $Z_G$ and of the cosmological constant,
they are just finite parts of the mode integral.

The divergent part of the trace $\calt=T^\mu_\mu$ is obtained from
the one of $\calf_{\rm div}$ by multiplying with 
$m(\tau) a^{-4+\epsilon}$ as
\be
\calt_{\rm fl, div}=a^
{-4+\epsilon}\left[4m^4 L_0+\frac{m_0^2m^2}{4\pi^2}
+2 m m''L_0 \right]
\pkt
\ee
Using the same procedure as above, and inserting all
the divergent counter terms as they have already been determined
we find the renormalized trace
\bea\nonumber
\calt_{\rm ren}&=&
4\calv_{\rm ren}(\tilde\phi)
\\&& +6\left(\xi-\frac{1}{6}\right)
a^{-2} \left(a^{-1}\tilde\phi'-H\tilde\phi \right)^2 
\\ &&\nonumber
+6(\zeta+\Delta \zeta)a^{-3}
\left(\tilde\phi''+\frac{2}{3}Ra^{-2}\tilde\phi\right)
+\frac{\Delta Z_{\rm G}}{\kappa}G^\mu_\mu
+4\frac{\Delta\Lambda}{\kappa}
\\&&\nonumber +\frac{ a^{-4}}{48\pi^2}
\left(a^2Rm^2-8a^2H^2m^2+12m_0^2m^2+4mm''\right.\\
&&\nonumber \left.+6m^4-2m^{'2}+16aHmm'\right)\\ \nonumber
&&+m\left(\calf_{\rm sing}+\calf_{\rm fin}\right)
\pkt\eea 
Thus far we have discussed the way of handling
the infinite terms by renormalization of the bare parameters.
We also have considered carefully the finite
terms left over after renormalization, in particular those that appear
due to the continuation of the space-time dimension via
divergent factors $1/(n-4)$ multiplying terms that vanish 
as $n-4$ as $n\to 4$.
We have considered, however, only a restricted class of metrics,
conformally flat ones. Furthermore, while our regularization is
relativistically covariant it is not so under general coordinate
transformations. As a consequence, the anomalous contributions
do not appear (they neither do, for the same reasons, 
in Ref. \cite{MaMo1} where a Pauli-Villars regularization is used). 
Indeed for the Dirac field we seemingly need no counter 
terms of the form $\,^{(1)}H_{\mu\nu}$, $\,^{(2)}H_{\mu\nu}$ or
 $H_{\mu\nu}$.
In general, however, one expects the divergent part of
$T_{\mu\nu}$ for non-interacting fermions
 to be (see, e.g.,  \cite{Davies:1977,Bunch:1979,BiDa})
\be \label{gendiv}
T_{\mu\nu}^{\rm div}=
\left\{\frac{8m^4}{n(n-2)}g_{\mu\nu}-\frac{2 m^2 }{3(n-2)}
G_{\mu\nu}-\frac{1}{180}\left[\frac{7}{4}H_{\mu\nu}
+2\,^{(2)}H_{\mu\nu}-\frac{5}{4}\,^{(1)}H_{\mu\nu}\right]  \right\}L
\kma\ee
where $L$ has been defined in the previous section, Eq. 
\eqn{Ldef}.
The first two terms in the parenthesis are cancelled by
the cosmological constant and gravitational wave function 
renormalization counter terms, Eq. \eqn{einsteinren}, they agree
with the divergences we find in our analysis.
The remaining terms are higher curvature terms, for whom
we have not found any divergences. This is due to the fact
that
for conformally flat metrics in $n=4$ one has the identity
\be
 H_{\mu\nu}=\,^{(2)}H_{\mu\nu}=\frac{1}{3}\,^{(1)}H_{\mu\nu}
\ee
and the expression in square brackets, i.e., the entire
higher curvature contribution vanishes, in agreement with
our analysis of divergent terms.
However, we have to supply
counter terms for those contributions as well,
as for metrics that are not conformally flat such
counter terms would be required. That means we have to subtract
these terms. Even in $n=4$ they yield a finite
contribution, the anomaly, if one continues the
expressions for those tensors from $n\neq 4$ to $n=4$
and takes into account the factor $2/\epsilon=2/(n-4)$
in the divergent factor $L$. The explicit expressions 
for these tensors for $n\neq 4$ are given in  Appendix A.
The finite contributions to $T_{\mu\nu}$ which we call
$T^{\rm an}_{\mu\nu}$ have been evaluated on the basis
of  these formulae
as well as \eqn{gendiv}, using {\sc maple}.
We find
\bea
T_{00}^{\rm an}
&=&\frac{1}{960\pi^2}\left\{
\frac{1}{12}R^2-H^2 R -\frac{11}{2}H^4-H\dot R\right\}
\\
T_{\mu}^{\mu\rm an}
&=&\frac{1}{960\pi^2}
\left\{-3H\dot R -\frac{11}{3}H^2 R +22
H^4 -\ddot R  \right\}
\pkt
\eea
The anomalous part of the energy-momentum tensor is conserved
separately from the nonanomalous part.


\section{The initial singularity}
\setcounter{equation}{0}

The fluctuation integral, the energy density, and the 
trace of the energy-momentum tensor
contain  contributions $\calf_{\rm sing},\cale_{\rm sing}$ and
$\calt_{\rm sing}$ that become singular as $\tau \to 0$. We have 
discussed previously \cite{BHPferm} how to get rid of such 
singularities by a Bogoliubov transformation. There we had assumed that
$\dot m(0)=0$. Here $m'(0)$ is necessarily different from zero,
as $a'(0)=H(0)a(0)^2$ is different from zero due to the first
Friedmann equation. So the singular terms are different and the singularities
are stronger. For the Minkowski geometry this has been discussed in
\cite{BBdVferm}.
From Eq. \eqn{fsing} we have,
using explicit expressions given in Appendix B of Ref. \cite{BHPinit},
\be
\calf_{\rm sing}(\tau) \simeq \frac{m'(0)}{4\pi^2}\tau ^{-1} 
+ \frac{m''(0)}{4\pi^2}\ln \tau
\pkt
\ee
The energy density stays finite as $\tau \to 0$ and the trace behaves as
\be
\calt_{\rm sing}(\tau) 
\simeq m(\tau)\left\{\frac{m'(0)}{4\pi^2}\tau^{-1}\ln \tau  + 
\frac{m''(0)}{4\pi^2}\ln \tau\right\}
\pkt
\ee
As it is simply proportional to the fluctuation integral,
it is sufficient to render the fluctuation integral finite.
To do so one performs a  Bogoliubov transformation
\bea
b_{\bfp,s}&=&\cos (\beta_{\bfp,s}) \tilde b_{\bfp,s}+
\sin(\beta_{\bfp,s})\exp^{i\delta_{\bfp,s}}\tilde d^\dagger_{-\bfp,s}
\\
d^\dagger_{-\bfp,s}&=&
-\sin(\beta_{\bfp,s})e^{-i\delta_{\bfp,s}}\tilde b_{\bfp,s}
+\cos (\beta_{\bfp,s}) \tilde d^\dagger_{-\bfp,s}
\eea
and defines the modified initial state as being annihilated 
by the new operators
$\tilde b_{\bfp,s}$ and $\tilde d^\dagger_{-\bfp,s}$.
The fluctuation integral then takes the form 
\bea
\tilde \calf(\tau)&=&\sum_s\inte\left\{\bar U_{\bfp, s} 
U_{\bfp,s}\sin^2\beta_{\bfp,s}
+\bar V_{-\bfp,s}V_{\-\bfp,s}\cos^2\beta_{\bfp,s}
\right.\nonumber\\
&&
\hspace{10mm}+\bar V_{-\bfp,s}U_{\bfp,s}e^{i\delta_{\bfp,s}}
\cos\beta_{\bfp,s}\sin\beta_{\bfp,s}\\
\nonumber 
&&
\hspace{10mm}\left.+\bar U_{\bfp,s} 
V_{-\bfp,s}e^{-i\delta_{\bfp,s}}\cos \beta_{\bfp,s}
\sin \beta_{\bfp,s}\right\}
\pkt\eea
or, explicitly
\bea \nonumber
\tilde\calf(\tau)
&=&-2 \sum_s\intn{n-1}{p}\left\{\cos(2\beta_{\bfp,s})
\left[\frac{m(\tau)}{E_0}
-\frac{ m'' (\tau)}{4(E_0)^3}-\frac{m^3(\tau)}{2(E_0)^3}+
\frac{m(\tau)m_0^2}{2(E_0)^3} \right.\right.
\\ &&\left. \label{ftildex}
-\frac{ m'(0)}{2(E_0)^2}
\sin(2E_0\tau) +
\frac{m''(0)}{4(E_0)^3}\cos (2E_0\tau)+K_F(p,\tau)
\right] 
\\ &&\left. \nonumber
+\sin(2\beta_{\bfp,s})\frac{ps}{2E_0(E_0+m_0)}
\left[-2 \im \left(f_pf_p'e^{i\delta_{\bfp,s}}\right)
+ 2 m \re \left(f_p^2 e^{i\delta_{\bfp,s}}\right)\right]
\right\}
\eea
Here we have used Eq. \eqn{flucintexp} and $\bar U_{\bfp,s} U_{\bfp,s}
=-\bar V_{\bfp,s} V_{\bfp,s}$
as well as
\be
\bar V_{-\bfp,s}U_{\bfp,s}=\left(\bar U_{\bfp,s}V_{-\bfp,s}\right)^*
= \frac{2ps}{E_0+m_0}\left(  i f_pf'_p +m (\tau)f_p^2\right)
\pkt
\ee  
In determining the parameters $\beta_{\bfp,s}$ and
$\delta_{\bfp,s}$ 
the mode functions in the last bracket of Eq \eqn{ftildex}
can be replaced by their leading behavior
\bea
2\im \left(f_pf_p'e^{i\delta_{\bfp,s}}\right)&\simeq& 2 E_0 
\cos(2E_0\tau -\delta_{\bfp,s})\\
2m(\tau)\re \left(f_p^2e^{i\delta_{\bfp,s}}\right)&\simeq& 
2m(\tau)\cos(2E_0\tau-\delta_{\bfp,s})
\pkt \eea
As we discuss the limiting behavior as $\tau\to 0$ we can 
replace $m(\tau)$ by $m_0$.
We then obtain the following conditions for the cancellation of the
singular contributions
\bea
-\cos(2\beta_{\bfp,s})\frac{m''(0)}{2E_0^2}+
2ps \sin 2\beta_{\bfp,s}\cos(\delta_{\bfp,s})=0
\\
\cos(2\beta_{\bfp,s})\frac{m'(0)}{E_0}+
2ps \sin 2\beta_{\bfp,s}\sin(\delta_{\bfp,s})=0
\eea
so that the parameters are obtained as
 \footnote{The right hand side of
Eq. (V.9) in \cite{BHPferm} should have a factor $4$ in the denominator,
instead of $8$, in accordance with Eq. (V.10).}
\bea
\tan \delta_{\bfp,s}&=& - 2 E_0 \frac{m'(0)}{m''(0)}
\\
\tan 2\beta_{\bfp,s}&=&
\frac{\ddot{m}(0)}{4sp(E_0)^2}\sqrt{1+\tan^2\delta}
\pkt
\eea
Here we have  retained only the leading asymptotic behavior. 
The resulting modifications of the various mode function integrals 
can be read off from those of Ref. \cite{BHPferm}.
One should note that $m''(0)$ does depend, via the
equation of motion for $\phi(\tau)$, on the 
modified fluctuation
integral $\calf(0)$ which now does not vanish as $\tau\to 0$,
and this again depends on $m''(0)$.
So $m''(0)$ has to be determined self-consistently.

The situation {\em before} removing the initial singularity is actually
even worse than it is apparent in the analysis of $\calf_{\rm sing}$.
By the second Friedmann equation a singularity of $\calt(\tau)$ entails
a singular behaviour of $a''(\tau)$, but $m(\tau)$ contains 
a contribution $a''(\tau)m_f$, so $m''(0)$ does not even exist.
This just reflects the impossibility of starting the evolution consistently
without getting rid, beforehand, of the initial singularity. 
An alternative way of avoiding these initial singularities has been discussed
recently for scalar fields, using adiabatic subtraction \cite{HaMoMo}.


\section{Conclusions}
\setcounter{equation}{0}

We have derived here the renormalized equations of motion for 
a massive Dirac field, with scalar coupling to a Yukawa field,
 in a flat FRW universe. We have used a formalism -
essentially a resolvent expansion -  that is useful
for numerical computations and allows at the same time for a straightforward
analysis of the divergent terms and their separation
from the finite parts to be computed numerically. 
Straightforward here does not mean
simple. For a massive Dirac field the Feynman diagrams for the
effective action are divergent
up to diagrams with $4$ external scalar fields. This 
reflects itself in a rather lengthy analysis, in our formalism as well.
We have not reproduced all the details here, but rely 
on our previous work \cite{BHPferm} for a Yukawa theory in
Minkowski space.
 
 The divergent terms obtained
in our formalism have been found in consistency
 with standard results, and the renormalization
counter terms can be chosen independent of the initial conditions.
The fact that we have considered a massive fermion field leads to 
a larger number of couplings of the scalar field,
necessary for renormalization. As these may be obtained
by a shift of the scalar field, this modification is
a straightforward generalization.
The renormalization counter terms have been determined in
dimensional regularization and with a $\overline { MS}$ convention
for the renormalization. The formalism can be adapted
to other regularization and renormalization schemes. 

The trace anomaly does not appear explicitly in our formalism.
This is a consequence of the special geometry we consider here and
does not imply an inconsistency of the formalism. 
This can be understood from the general analysis of the divergent parts
of the energy-momentum tensor, as given, e.g., in \cite{BiDa}.
We have used this analysis in order to infer the anomalous part of the
energy-momentum tensor.

We have also analyzed the initial singularities. They obstruct 
the cosmological initial value problem in an essential way,
 a singularity in the
trace of the energy momentum tensor leads 
to a singularity in the expansion rate. So they have to be avoided by
a suitable choice of the initial state. We have constructed
explicitly such a state; it is not unique but can be modified
if only the asymptotic behavior as $p\to\infty$ of the Bogoliubov
parameters is retained.

Our results can be used in a straightforward way for numerical 
computations, which we plan to carry out in the near future. 
We do not expect an essential modification of results
related to the parametric resonance phenomenon; however, especially
for large Yukawa couplings and/or masses, the finite remnants 
of the renormalization
counter terms and the singular parts are not necessarily small 
and this signals the importance of considering renormalization
carefully.

\section{Acknowledgments}
It is a pleasure to thank
D. Boyanovsky, B. Bassett, H. de Vega, P. B. Greene, K. Heitmann, L. Kofman,
A. Mazumdar, and A. Riotto for useful and inspiring discussions.

\vspace*{20mm}

\begin{appendix}

\section{The higher curvature tensors}
\setcounter{equation}{0}
The higher curvature tensors are defined by
\bea
^{(1)}H\,_{\mu\nu}&&=\frac{1}{\sqrt{-g}}\frac{\delta}{\delta g\,^{\mu\nu}}
\int\!d^nx \sqrt{-g}\,R^2\nonumber\\
&&=2\,R_{;\mu\nu}-2g\,_{\mu\nu}\;\Box R
-\frac 1 2 g\,_{\mu\nu}\,R^2+2RR\,_{\mu\nu}\ko\\
^{(2)}H\,_{\mu\nu}&&=\frac{1}{\sqrt{-g}}\frac{\delta}{\delta g\,^{\mu\nu}}
\int\!{d}^nx \sqrt{-g}\,R\,^{\alpha\beta}R\,_{\alpha\beta}\nonumber\\
&&=2 R\,^{\alpha}_{\mu\, ;\nu\alpha}- 
\Box R_{\mu\nu}-\frac 1 2g\,_{\mu\nu}\,\Box R
+2R\,_{\mu}^{\alpha}R\,_{\alpha\nu}-
\frac{1}{2}g\,_{\mu\nu}\,R\,^{\alpha\beta}
R\,_{\alpha\beta}\nonumber\\
H\,_{\mu\nu}&&=\frac{1}{\sqrt{-g}}\frac{\delta}{\delta g\,^{\mu\nu}}
\int\!{d}^nx \sqrt{-g}\,R\,^{\alpha\beta\gamma\delta}
R\,_{\alpha\beta\gamma\delta}\nonumber\\
&&=
-\frac 1 2 g\,_{\mu\nu}R\,^{\alpha\beta\gamma\delta}
R_{\alpha\beta\gamma\delta}+2
R\,^{\mu\alpha\beta\nu}R\,_{\nu}^{\alpha\beta\gamma}
-4\Box R\,_{\mu\nu}+2R\,_{;\mu\nu}\nonumber\\
&&
-4R\,_{\mu\alpha}R\,^{\alpha}_{\nu}+4R\,^{\alpha\beta}
R\,_{\alpha\mu\beta\nu}\pkt
\eea
On order to evaluate the anomalous contribution to the
energy-momentum tensor we need the continuation
of the higher order curvature tensors for the FRW geometry
 to $n=4-\epsilon$ dimensions. We choose 
to continue the spatial dimension to $3-\epsilon$.
We already have given the $n$-dimensional continuation
of the Ricci tensor in Eq. \eqn{Ricci_ndim}.

For the time-time components and the trace of the tensors 
$\,^{(n)}H_{\mu\nu}$ we obtain
\bea
\label{htt1}
\,^{(1)}H\,_{tt}&=&-6H\dot R +\frac 1 2 R^2- 6 H^2R
\\ \nonumber &&+ (n-4)
\left(-2H\dot R-(n+1)RH^2\right)\kma\\ 
\label{htt2}
^{(2)}H\,_{tt}&=&-2H\dot R+\frac 1 6 R^2-2H^2R+(n-4)\left(-\frac 1 2 H\dot R
\right.\\ \nonumber
&&\left.-\frac{R^2}{24(n-1)}
-\frac 1 4(n+2)H^2R+\frac 1 8 (n-1)(n-2)^2H^4\right)\kma\\
\label{htt3}
H\,_{tt}&=&-2H\dot R+\frac 1 6 R^2-2H^2R\\ \nonumber
&&+(n-4)\left(
-\frac{R^2}{6(n-1)}-H^2R+\frac 1 2(n-1)(n-2)H^4
\right)\kma
\eea
\bea \label{hmumu1}
^{(1)}H\,_{\mu}^{\mu}&=&-6\ddot R-18H\dot R+(n-4)\left(
-2\ddot R-2(n+2)H\dot R-\frac 1 2 R^2\right)\kma\\
\label{hmumu2}
^{(2)}H\,_{\mu}^{\mu}&=&-2\ddot R -6 H\dot R +(n-4)
\left(-\frac 1 2 \ddot R-\frac 1 2 (n+3)H\dot R
\right.\\ \nonumber
&&\left.-\frac{nR^2}{8(n-1)}+\frac 1 4 
(n-2)^2H^2R-\frac 1 8n(n-1)(n-2)^2H^4\right)\kma\\
\label{hmumu3}
H\,_{\mu}^{\mu}&=&-2\ddot R -6H\dot R +(n-4)
\left(-2H\dot R -\frac{R^2}{2(n-1)}
\right.\nonumber\\&&\left.+(n-2)H^2R-\frac 1 2 n (n-1)(n-2)H^4\right)
\pkt
\eea


\section{The Dirac equation in $n$ dimensional FRW geometry}
\setcounter{equation}{0}
The Dirac equation in curved space is formulated using the
$n$-bein $e^a=e^a_\mu dx^\mu$ and the  
spin connection $\omega^{ab}=\omega^{ab}_\nu dx^\nu$.
 It has the form
\be \label{dirac_general}
i\gamma^a E_a^\nu\left(\partial_\nu+\omega^{ab}_\nu
\frac{1}{8}\left[\gamma_a,\gamma_b\right]\right) \psi
-m \psi =0
\pkt \ee
Here $E_a^\mu$ is the inverse $n$-Bein and the connection is
defined by the equation
\be
d\wedge e^a + \omega^a_b\wedge e^b =0
\ee
Choosing for the FRW geometry $e^0=dt$ and $e^k=a\, dx^k$
one finds
\be
{\omega^k}_0 = \dot a \, dx^k 
\kma
\ee
while the purely spatial components ${\omega^k}_j$ vanish.
Using  $E^0_0=1$, $E^k_k=1/a$, all other components being zero,
we obtain
\be
i \left(\gamma^0 \partial_0 +\frac{1}{a}\gamma^k\nabla_k
+ \frac{1}{a}\gamma^k{\omega^{k0}}_k\frac{1}{4}
\left[\gamma_k,\gamma_0\right]\right)\psi-m\psi=0
\kma\ee
or, using $\gamma^k\left[\gamma_k,\gamma_0\right]=2(n-1)\gamma^0$
\be
i\gamma^0\left(\partial_0+\frac{n-1}{2}\frac{\dot a}{a}\right)\psi
+ \left(\frac{i}{a}\gamma^k\nabla_k -m\right)\psi=0
\pkt
\ee
It is the almost obvious generalization of the Dirac equation
 in the  $n=4$ FRW geometry.
 
\end{appendix}



\begin{thebibliography}{10}

\bibitem{Inflation} see, e.g.,
 L. F. Abbott and S.-Y. Pi,
{\em Inflationary Cosmology} (World Scientific, Singapore 1986);
E.~W. Kolb and M.~S. Turner, {\em The early universe} (Addison Wesley,
Redwood City, 1990), A. Linde, {\em Particle physics and inflationary cosmology}
(Harwood Academic Publishers, Chur, 1990).

\bibitem{KoLiSta}
L. Kofman, A. Linde, and A.~A. Starobinskii, Phys. Rev. Lett.
{\bf 73}, 3195 (1994); Phys. Rev {\bf D56}, 6175 (1997).

\bibitem{ShtTrBr}
Y. Shtanov, J. Traschen, and R. Brandenberger, Phys. Rev. {\bf D51},
  5438  (1995).

\bibitem{BodVHo}
D. Boyanovsky, H.~J. de~Vega, and R. Holman, Phys. Rev. {\bf D49},  2769
  (1994).

\bibitem{BoCodVHo}
D. Boyanovsky, D. Cormier, H.~J. de~Vega, and R. Holman, 
Phys. Rev. {\bf D55}, 3373  (1997).

\bibitem{BoCodVHiSiSre}
D. Boyanovski, D. Cormier, H. J. de Vega, R. Holman,
A. Singh, and M. Srednicki, Phys.Rev. {\em D56}, 1939 (1997). 

\bibitem{BoCodVHoKu}
D. Boyanovsky, D. Cormier, H.J. de Vega, R. Holman, 
and S.P. Kumar,  Phys. Rev. {\bf D57}, 2166 (1998) 

\bibitem{RaHu}
S.A. Ramsey and B.L. Hu, 
Phys. Rev. {\bf D56} 678 (1997). 

\bibitem{BHPfrw}
J. Baacke, K. Heitmann, and C. P\"atzold, 
Phys.  Rev. {\bf D56}, 6556 (1997).

\bibitem{BPfrw} J. Baacke and C. P\"atzold, 
{\em Out of equilibrium evolution of scalar fields
in FRW cosmology: renormalization and numerical simulations},
hep-ph/9906417, Phys. Rev. {\bf D}, to appear. 

\bibitem{BHPnumrec}
J. Baacke, K. Heitmann, and C. P\"atzold, 
Phys.  Rev. {\bf D55}, 2320 (1997).


\bibitem{Tsujikawa:1999a}
S.~Tsujikawa, K.~Maeda and T.~Torii,
Phys.\ Rev.\  {\bf D60} (1999) 063515;
{\em ibid.} 123505;

\bibitem{Tsujikawa:1999b}
S.~Tsujikawa, K.~Maeda and T.~Torii,
{\em Preheating of the nonminimally coupled inflaton field},
hep-ph/9910214.



\bibitem{Molina-Paris:1999}
C.~Molina-Paris, P.~R.~Anderson and S.~A.~Ramsey,
{\em One-loop lambda phi**4 field theory in Robertson-Walker 
spacetimes:  Adiabatic regularization and analytic approximations},
gr-qc/9908037.

\bibitem{HaMoMo}
S.~Habib, C.~Molina-Paris and E.~Mottola,
{\em Energy-momentum tensor of particles created in an expanding universe},
Phys.\ Rev.\  {\bf D61} (2000) 024010.


\bibitem{BdVHSal}
D. Boyanovsky, H. J. de Vega, R. Holman, and J. F. J. Salgado,
Phys. Rev. {\bf D54}, 7570 (1996). 

\bibitem{Sal}
J. Salgado, {\em Hierarchy of sum rules in out-of-equilibrium
quantum field theory}, DO-TH-99-06, May 1999; hep-th/9905106. 

\bibitem{BHPferm}
J. Baacke, K. Heitmann, and C. P\"atzold, 
Phys. Rev. {\bf D58}, 125013 (1998).

\bibitem{GreeKo}
P. B. Greene and L. Kofman, Phys.Lett. {\bf B448},6 (1999). 

\bibitem{MaMa1}
A. L. Maroto and A. Mazumdar, Phys. Rev. D 59, 083510 (1999).
 
\bibitem{ChuKoRiTka}
D. J. H. Chung, E.W. Kolb, A. Riotto, I. Tkachev, 
{\em Inflation and features in the primordial power spectrum},
hep-ph/9910437.

\bibitem{MaMa2}
A. L. Maroto and A. Mazumdar,
{\em Production of spin 3/2 particles from vacuum fluctuations},
hep-ph/9904206.

\bibitem{GiuRiTka}
G.F. Giudice, A. Riotto, and I. Tkachev,
JHEP 9911:036 (1999). 

\bibitem{KaKoLivPr}
R. Kallosh, L. Kofman, A. Linde, and A. van Proeyen,
{\em Gravitino production after inflation}, hep-th/9907124.

\bibitem{GiuTkaRi}
G.F. Giudice, I. Tkachev, and A. Riotto,
JHEP 9908:009 (1999).

\bibitem{GiuPeRiTka}
G.F. Giudice, M. Peloso, A. Riotto, and I. Tkachev,
JHEP 9908:014 (1999).

\bibitem{PaFu}
L. Parker and S. A. Fulling, Phys. Rev. {\bf D9}, 341 (1984).

\bibitem{Davies:1977} P. C. W. Davies, S. A. Fulling, S. M. 
Christensen, and T. S. Bunch, Ann. Phys. (N.Y.) {\bf109}, 108 (1977).
 
\bibitem{Bunch:1979} T. S. Bunch, J. Phys. A:
Math. Gen. {\bf 12}, 517 (1979); {\em ibid.} {\bf 13}, 1297 (1980).

\bibitem{BiDa}
N.~D. Birrell and P.~C.~W. Davies, {\em Quantum fields in curved space},
Cambridge University Press,   Cambridge, England, 1982.

\bibitem{MaMo1}
S.G. Mamaev and V.M. Mostepanenko,  
 Yad.Fiz. {\em 27} ,1640 (1978). 

\bibitem{MaMo2}
S.G. Mamaev and V.M. Mostepanenko,  
 Yad.Fiz. {\em 37} ,1323 (1983).

\bibitem{CooMoinit} F. Cooper and E. Mottola,
 Mod. Phys. Lett. 2, 635
(1987);  Phys. Rev. {\bf D36}, 3114 (1987). 

\bibitem{BHPinit}
J. Baacke, K. Heitmann, and C. P\"atzold, 
Phys.  Rev. {\bf D57}, 6398 (1998).

\bibitem{BBdVferm}
J. Baacke, D. Boyanovsky, and H. de Vega,
{\em Initial time singularities in nonequilibrium evolution of condensates
and their resolution in the linearized approximation},
hep-ph/9907337.

\end{thebibliography}
\end{document}